\documentclass[showkeys,aps,amsmath,amssymb,twocolumn,groupedaddress,longbibliography,floats,final,postprint,superscriptaddress]{revtex4-1}
\usepackage{enumerate}
\usepackage{graphicx}
\usepackage{amsmath,amssymb,amsthm,enumitem}

\usepackage{color}
\definecolor{LinkColor}{RGB}{199,21,133}
\usepackage[colorlinks=true,linkcolor=blue,anchorcolor=blue,citecolor=blue,urlcolor=blue]{hyperref}
\usepackage[polutonikogreek,english]{babel}

\usepackage{hyperref}

\usepackage{epstopdf}
\usepackage{type1cm}
\usepackage{multirow}
\usepackage{times}

\newcommand{\scrN}{{\mathcal N}}
\newcommand{\scrS}{{\mathcal S}}

\newcommand{\scrO}{{\mathcal O}}
\newcommand{\scrR}{{\mathcal R}}
\newcommand{\scrC}{{\mathcal C}}
\newcommand{\scri}{{\mathcal I}}

\begin{document}

\title{History-dependent percolation in two dimensions}

\author{Minghui Hu}
\author{Yanan Sun}
\author{Dali Wang}
\author{Jian-Ping Lv}
\email{jplv2014@ahnu.edu.cn}
\affiliation{Department of Physics, Anhui Normal University, Wuhu, Anhui 241000, China}
\author{Youjin Deng}
\affiliation{National Laboratory for Physical Sciences at Microscale and Department of Modern Physics, University of Science and Technology of China, Hefei, Anhui 230026, China}
\affiliation{MinJiang Collaborative Center for Theoretical Physics, Department of Physics and Electronic Information Engineering, Minjiang University, Fuzhou, Fujian 350108, China}
\date{\today}
\begin{abstract}
We study the history-dependent percolation in two dimensions, which evolves in generations from standard bond-percolation configurations through iteratively removing occupied bonds. Extensive simulations are performed for various generations on periodic square lattices up to side length $L=4096$. From finite-size scaling, we find that the model undergoes a continuous phase transition, which, for any finite number of generations, falls into the universality of standard 2D percolation. At the limit of infinite generation, we determine the correlation-length exponent $1/\nu=0.828(5)$ and the fractal dimension $d_{\rm f}=1.864\,4(7)$, which are not equal to $1/\nu=3/4$ and $d_{\rm f}=91/48$ for 2D percolation. Hence, the transition in the infinite-generation limit falls outside the standard percolation universality and differs from the discontinuous transition of history-dependent percolation on random networks. Further, a crossover phenomenon is observed between the two universalities in infinite and finite generations.
\end{abstract}

\maketitle
\section{Introduction}\label{int}
Percolation, originally proposed for modeling transport behavior in a random medium~\cite{broadbent1957percolation}, has numerous applications in various areas of science and technology~\cite{grimmett1999percolation}. In the standard bond percolation on a given lattice, each bond is independently occupied with probability $p$, and a cluster corresponds to a set of sites connected together by occupied bonds. As $p$ increases, the bond percolation undergoes a continuous transition at percolation threshold $p_c$ from a state of locally connected sites to the percolating phase with an infinitely spanning cluster~\cite{stauffer2018introduction}. In two dimensions,
the critical exponents $\nu=4/3$ (for correlation length) and $\beta/\nu=5/48$ (for order parameter) are predicted by conformal field theory~\cite{Cardy1987}, Coulomb gas theory~\cite{nienhuis1987two} and stochastic Loewner evolution~\cite{lawler2000dimension}, and confirmed exactly in the triangular-lattice site percolation~\cite{Smirnov2001}. In the renormalization-group treatment, these exponents are related to the thermal and magnetic renormalization exponents as $y_t=1/\nu=3/4$ and $y_h=2-\beta/\nu=91/48$. The magnetic exponent $y_h$ is also referred to as the fractal dimension $d_{\rm f}$ for critical percolation clusters.

Phase transitions in unconventional percolation models has been a compelling topic~\cite{araujo2014recent}. Explosive percolation transitions on random graphs~\cite{PhysRevLett.105.255701,PhysRevLett.103.255701,riordan2011explosive} and square lattices~\cite{radicchi2010explosive,PhysRevE.82.051105} were confirmed to be continuous. Rare examples of discontinuous percolation transition come from the bootstrap percolation~\cite{chalupa1979bootstrap,adler1991bootstrap,Choi2020} and cascading failure~\cite{buldyrev2010catastrophic,son2012percolation} models. Persistent attention has been paid to critical phenomena of the percolation models that involve manipulations of cluster structures~\cite{PhysRevLett.116.055701,hilario2019bernoulli,PhysRevE.50.R2403,liu2015recursive,li2020history}.

In real-world networks, there may simultaneously exist different types of connections between
pairs of sites (nodes), and these different connections can interact with each other.
For instance, information can spread on social networks with multiple communication channels.
It is common that various diseases coexist in society and the spread of a disease depends on the spread of other diseases and immunization information. The history-dependent percolation (HDP) model was proposed as a primitive model to mimic some basic features of such multiplex networks~\cite{li2020history}. It has been demonstrated~\cite{li2020history} that the history-dependent process can extract crucial characteristics of networks from the empirical data of brain scans and social networks.

\begin{figure*}
    \centering
    \includegraphics[width=14cm]{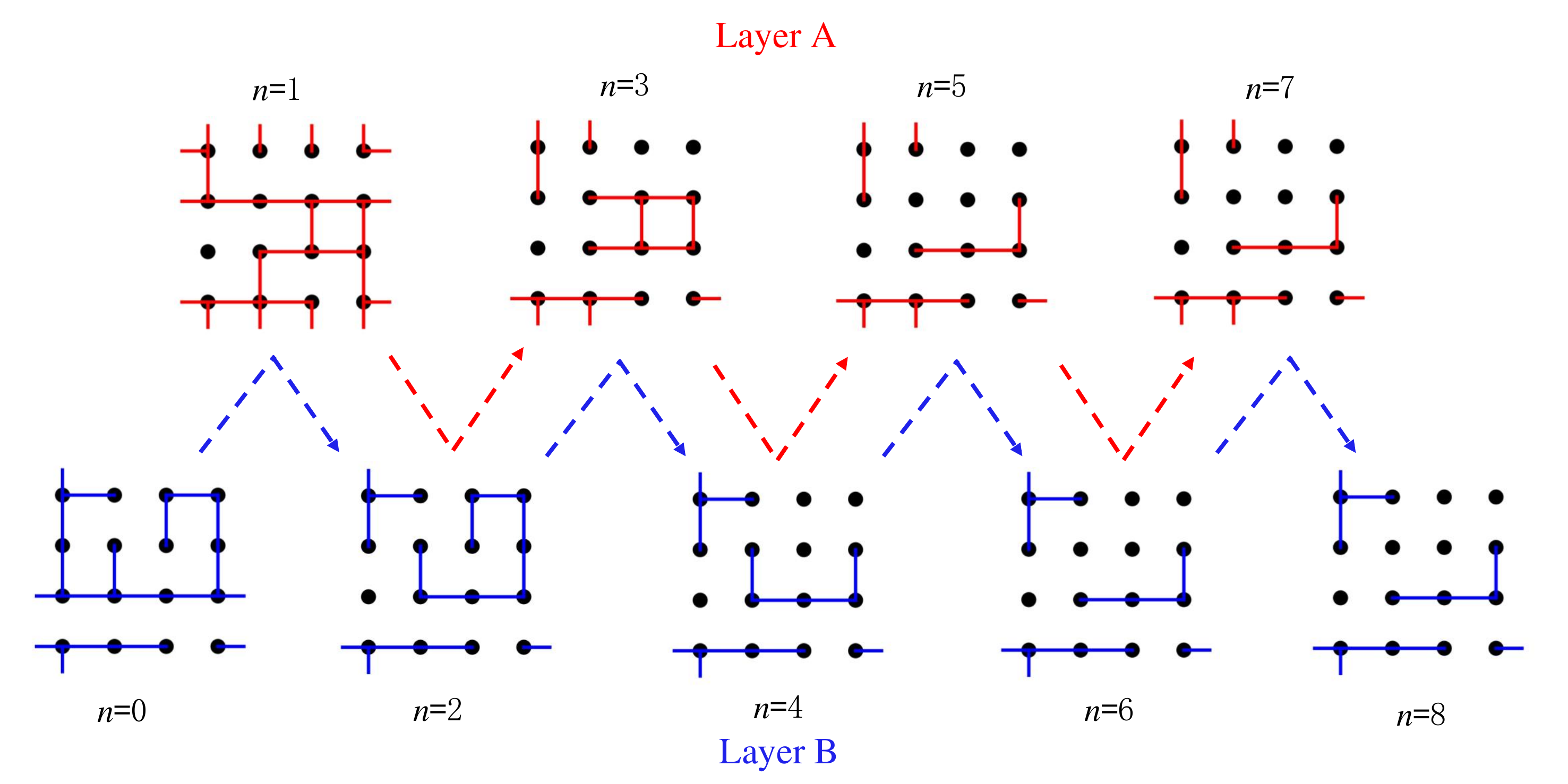}
  \caption{Illustration for the evolution in the HDP on a periodic square lattice. Generations $n=0$ and $n=1$ are independent configurations produced by a random process as of standard bond percolation. The generation $n$ with $n \geq 2$ evolves from the prior generation $n-2$ in the same layer, with the input from generation $n-1$ in the other layer: for each occupied bond in generation $n-2$, if its two ending sites belong to different clusters in generation $n-1$, remove the occupied bond. The process is halted once two consecutive evolutions do not make any change to configurations. For this specific example, the generation number after the last evolution is $\scri=8$.}~\label{fig_guide}
\end{figure*}

The HDP model has the bond-occupation probability $p$ as a free parameter (as in the standard bond percolation),
while it introduces the coupling between different types of connections in a dynamic process. An example is illustrated in Fig.~\ref{fig_guide}.
First, one generates two random and uncorrelated bond configurations with probability $p$,
labeled as generation $n \! = \! 0$ and $1$.
Then, one sequentially visits each occupied bond in the $n \! = \! 0$ configuration,
and deletes the occupied bond iff its two ending sites belong to different clusters in the $n \! = \! 1$ configuration.
Applying this operation to the $n\! = \!0$ generation deterministically leads to a new configuration, which is specified by $n \! = \! 2$.
Analogously, the $n \! = \! 3$ configuration is obtained from $n \! = \! 1$ according to its coupling to the $n \! = \! 2$ configuration.
Repeat the procedure until no more bonds can be removed--i.e., the configurations become saturated. For $n=0$ and $1$, the model is simply the standard bond percolation, and for $n \geq 2$, it is called HDP.

\begin{figure}
\centering
\includegraphics[width=8cm,height=6.5cm]{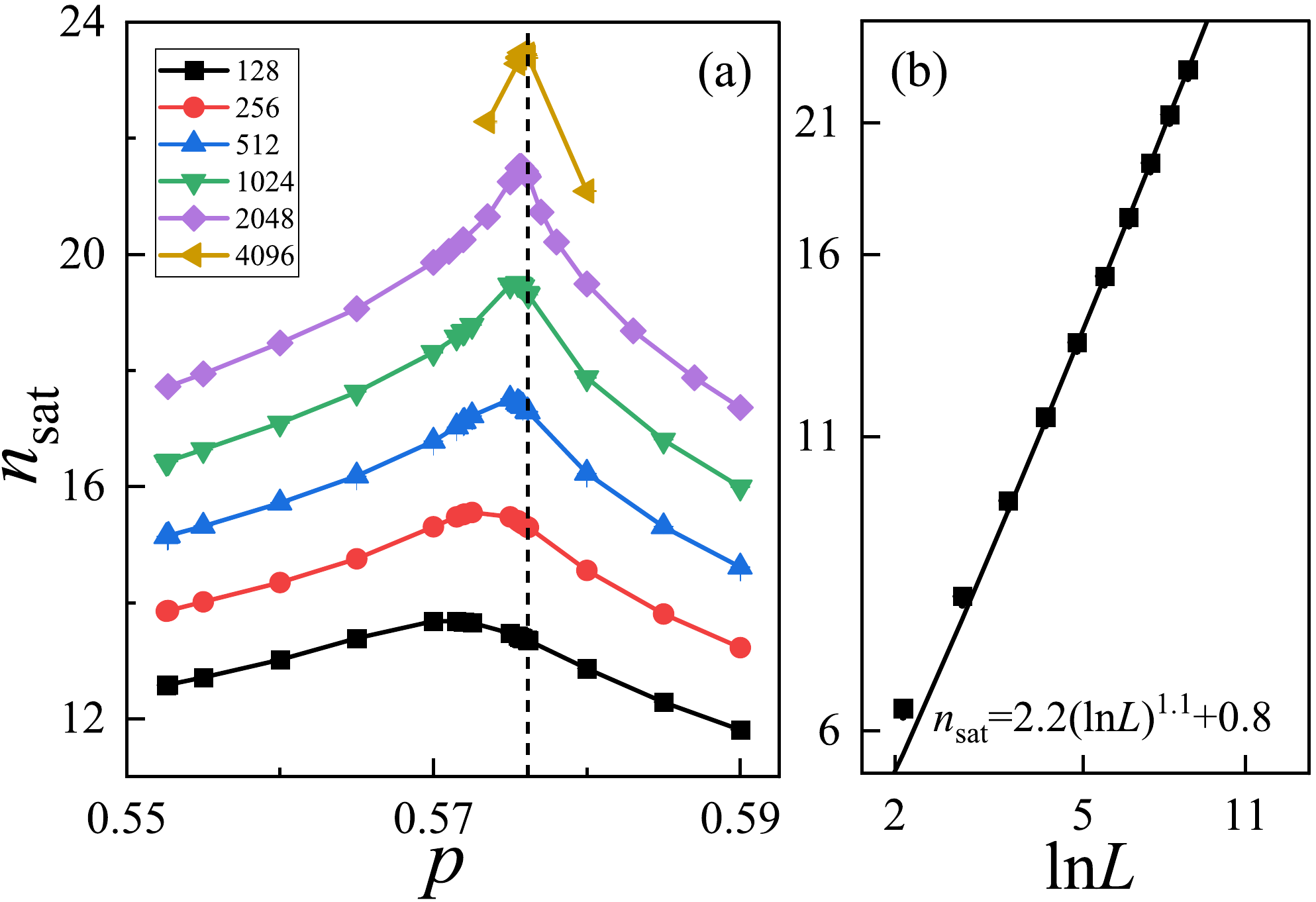}
\caption{Quantity $n_{\rm sat}$ versus $p$ for various $L$ (a) and versus ${\rm ln} L$ at $p=0.576\,132$ (percolation threshold for infinite generation) (b) on the square lattice. Panel (b) is plotted in a log-log scale. The dashed line in panel (a) marks the position $p=0.576\,132$ and the solid line in panel (b) represents the logarithmic divergence $n_{\rm sat}=2.2({\rm ln}L)^{1.1}+0.8$.}~\label{fig_np_n}
\end{figure}

The generation number $\scri$~\footnote{In actual simulations, we halt the iterations once two consecutive evolutions do not make any change to configurations, and the generation number of ending configuration is taken as $\scri$.} of saturated generation is a random number depending on the initial $n\! = \!0$ and 1 configurations. On the square lattice, Fig.~\ref{fig_np_n}(a) plots the Monte Carlo data for the statistical average $n_{\rm sat} \equiv \langle {\mathcal I} \rangle$ as a function of the initial bond-occupation probability $p$. For a given system size $L$, $n_{\rm sat}$ has a maximum at $p \approx 0.57$.  As $L$ increases, the peak location quickly converges to $p=0.576\,132$, which corresponds to the percolation threshold in the $n \rightarrow \infty$ limit (as shown later). Besides, there is an important feature that, irrespective of the bond-occupation probability $0 < p <1$, the $n_{\rm sat}$ value diverges approximately as $\ln L$, as illustrated in Fig.~\ref{fig_np_n}(b). This means that, given any finite value of $n$, the configuration at the $n$th generation is not saturated as long as $L$ is sufficiently large. Thus, to explore the HDP model in the $n \rightarrow \infty$ limit, one has to repeat the aforementioned evolution till the saturated generation, and sample quantities of interest from the corresponding saturated configurations.

In Ref.~\cite{li2020history}, the phase transition of HDP was studied as a function of $n$ on randomly networked structures including Erd\H{o}s--R\'{e}nyi network and scale-free network.
For any given {\it finite} generation $n \geq 2$, a percolation transition was found to be in the mean-field percolation universality class.
Directly in the $n \! \rightarrow \! \infty$ limit, for which the configurations in actual simulations are taken from those at $\scri$, it was shown that the percolation transition is discontinuous.

 Given that lattices and random networks are complementary testbeds for the insights into percolation transitions, we study the HDP on the square lattice. The remainder of the paper is organized as follows. Section~\ref{smf} summarizes main findings. Section~\ref{ms} introduces numerical methodology with an emphasis on sampled quantities. Section~\ref{rescs} presents numerical results: subsection~\ref{pt} and \ref{cpyy} present the determination of percolation thresholds and critical exponents, respectively; subsection~\ref{csci} demonstrates the geometric properties of critical clusters for infinite generation; subsection~\ref{ccifg} focuses on the crossover behavior of critical phenomenon from infinite to finite generation. Discussions are finally given in Sec.~\ref{sm}.

\section{Summary of main findings}\label{smf}
\begin{figure}
    \centering
    \includegraphics[width=8cm]{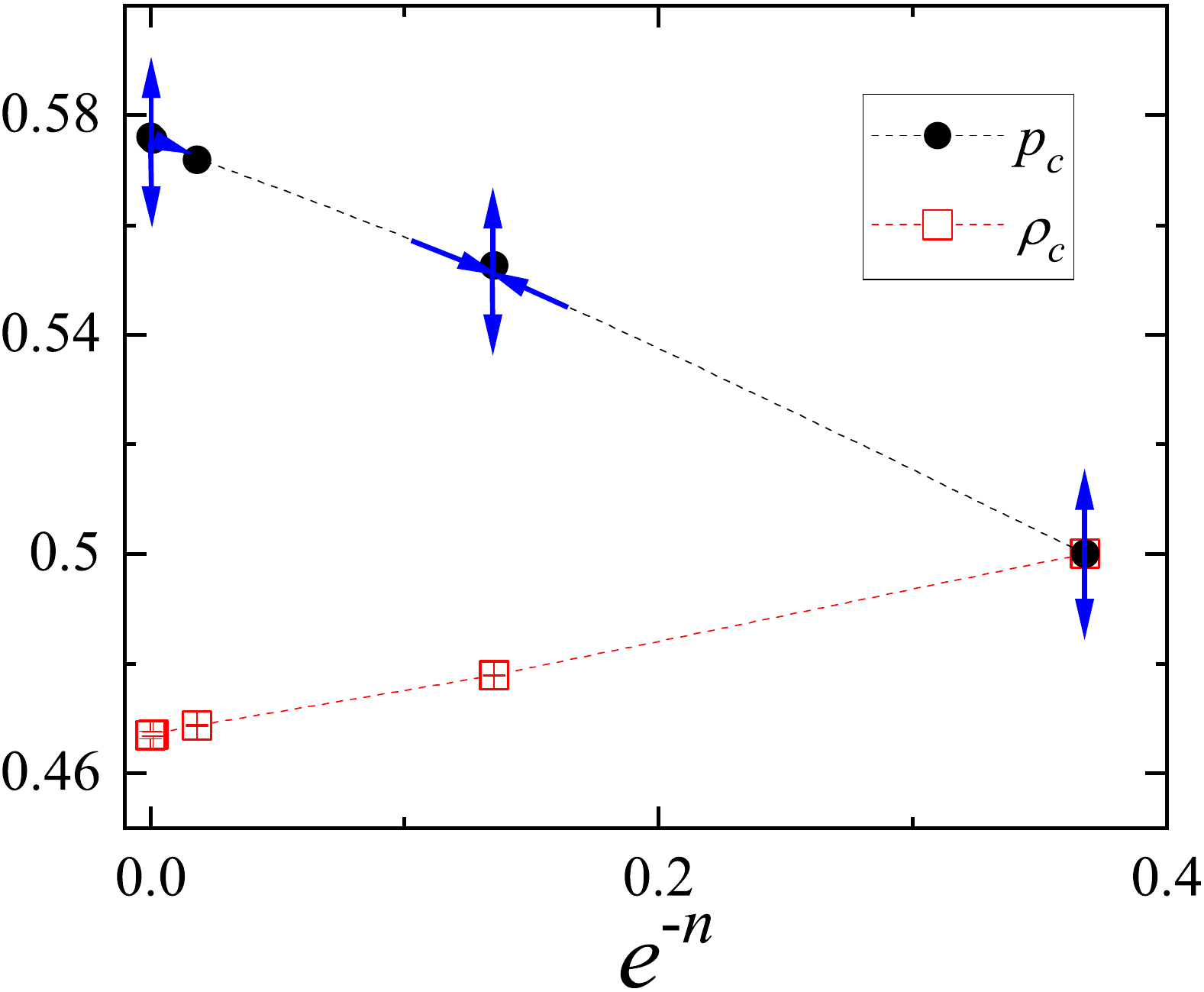}
  \caption{The dependence of percolation threshold $p_c$ and critical bond-occupation density $\rho_c$
  on $n$ for the HDP on the square lattice.
  For $n=1$, one has $p_c=\rho_c=1/2$.
   At a finite $n$, the continuous phase transition at $p_c$ has critical exponents
  consistent with $y_t=3/4$ and $y_h=91/48$. At the infinite generation $n \to \infty$, there is a continuous transition at $p_c$ with critical exponents $y_t=0.828(5)$ and $y_h=1.864\,4(7)$. Here, we plot $p_c$ and $\rho_c$ versus $e^{-n}$ to indicate their fast convergence as $n$ increases, in the sense that $p_c$ and $\rho_c$ versus $e^{-n}$ relations are close to linearities. Renormalization flows around fixed points are sketched by arrows.}~\label{pts}
\end{figure}

We perform Monte Carlo simulations for the HDP on periodic square lattices with $L$ ranging from $L=8$ to $4096$. For each generation $n \! \geq \! 2$, the connectivity of the corresponding graph is investigated as a function of the bond-occupation probability $p$,
which is used to generate the $n \! = \! 0$ and $1$ configurations, and the percolation threshold $p_c$ is determined. As shown in Fig.~\ref{pts}, $p_c$ increases with $n$.

 For the finite generations $n=2$, $4$ and $7$, we find that the model exhibits a continuous transition whose critical exponents are consistent with $y_t=3/4$ and $y_h=91/48$ for the standard uncorrelation percolation model in two dimensions. In the infinite-generation limit, a continuous phase transition is observed at $p_c = 0.576\,132 (2)$. However, the estimated thermal and magnetic renormalization exponents, $y_t=0.828(5)$ and $y_h=1.864\,4(7)$, are significantly different from those for the two-dimensional (2D) percolation model, indicating the emergence of a new universality class in the $n \rightarrow \infty$ limit.

The continuous transition in infinite generation is further confirmed from the critical distribution function $P(\scrC_1,L)$ of the largest-cluster size $\scrC_1$, which follows a single-variable function $\tilde{P}(x)$ with $P(\scrC_1,L)d\scrC_1=\tilde{P}(x)dx$ [$x \equiv \scrC_1/L^{d_f}$, $d_f=y_h=1.864\,4(7)$]. In addition, at $p_c$, the cluster-number density $n(s,L)$ of size $s$ obeys the standard scaling formula $n(s,L) \sim s^{-\tau} \tilde{n} (s/L^{d_f})$ of a continuous transition, with the hyper-scaling relation $\tau=1+2/d_f$.

To further explore the distinct universality classes for the infinite and the finite generation, we demonstrate the crossover phenomenon from the size-dependent behavior of various quantities and self-defined effective critical exponents.

\section{Methodology, quantities of interest, and scaling ansartz}\label{ms}

For each generation of the percolation configurations, we identify clusters of connected sites using the breadth-first search, and sample observables that are analogous to those in high-precision Monte Carlo studies of standard percolation models~\cite{PhysRevE.66.016129,wang2013bond,xu2014simultaneous} and relevant models~\cite{lv2012scaling,xu2019high,lv2020}.

In principle, the infinite generation stems from standard bond percolation ($n=0$) through an infinite number of generations. In practice, for $n \ge \scri$, the configurations on each layer (Fig.~\ref{fig_guide}) no longer change with increasing $n$ and are already in the infinite-generation limit. We further confirm the equivalence of sampling infinite generations on the two layers; in what follows, we analyze
the results of infinite generation with Layer B.

More specifically speaking, for each generation $n$, the following observables are defined:

\begin{itemize}
 \item The number of occupied bonds $\scrN$ remaining at the generation.
 \item The size $\scrC_1$ of the largest cluster.
 \item The second moment of cluster-size distribution $\scrS_2 = \sum_{C}|C|^2$, where the summation runs over all clusters.
 \item The observables $\scrR^{(x)}$ and $\scrR^{(y)}$, which equal $1$ if a cluster wraps around the periodic lattice in $x$ and $y$ direction, respectively, and equal $0$ otherwise.
\end{itemize}

For each generation, we sample the following quantities using aforementioned observables:
 \begin{itemize}
 \item The density of occupied bonds $\rho=\langle \scrN \rangle/(2L^2)$. For $n \! = \! 0$ and 1, one has $\rho=p$.
 \item The mean size of the largest cluster $C_1 = \langle \scrC_1 \rangle$.
 \item A susceptibility-like quantity $\chi=\langle \scrS_2 \rangle/L^2$.
 \item The wrapping probability
 \begin{equation}
  R^{(h)} = \langle \scrR^{(x)} \rangle = \langle \scrR^{(y)} \rangle,
 \end{equation}
 which gives the probability that a wrapping exists in the $x$ direction. In particular, for the 2D {\it standard} percolation, the critical value $R^{(h)}_{c}=0.521\,058\,290$ is exact in the $L \to \infty$ limit~\cite{newman2001fast}.

 \item Let $\scrN_{0,1}$ be the total number of occupied bonds in generation $0$ and $1$. We define the covariance of $\scrR^{(x)}$ and $\scrN_{0,1}$ as
 \begin{equation}
   g^{(h)}_{bR} = \langle \scrR^{(x)} \scrN_{0,1} \rangle - \langle \scrR^{(x)}\rangle \langle \scrN_{0,1} \rangle,
 \label{eq:g}
 \end{equation}
 which relates to the derivative of $R^{(h)}$ with respect to $p$ by $g^{(h)}_{bR} =p(1-p) \frac{dR^{(h)}}{dp}$.
\end{itemize}

For analyzing continuous phase transitions, we employ the tool of finite-size scaling (FSS) theory, which predicts that a quantity $Q$ near criticality scales as
\begin{eqnarray}\label{scalingQs}
Q(L,p)=L^{X_Q} \widetilde{Q}((p-p_c)L^{y_t}),
\end{eqnarray}
where $\widetilde{Q}$ is a scaling function. The scaling exponent $X_Q$ is quantity-dependent.  Quantities $C_1$ and $\chi$ can be related to order parameter and susceptibility, and hence have scaling exponents $y_h$ and $2y_h-2$, respectively~\cite{stauffer2018introduction}. Quantity $R^{(h)}$ is dimensionless ($X_Q=0$)~\cite{newman2001fast}, while the scaling exponent for $g^{(h)}_{bR}$ is $y_t$~\cite{wang2013bond,xu2014simultaneous}.

\section{Numerical results}\label{rescs}
\subsection{Percolation thresholds}\label{pt}
\begin{figure}
\centering
\includegraphics[width=8.8cm,height=7.8cm]{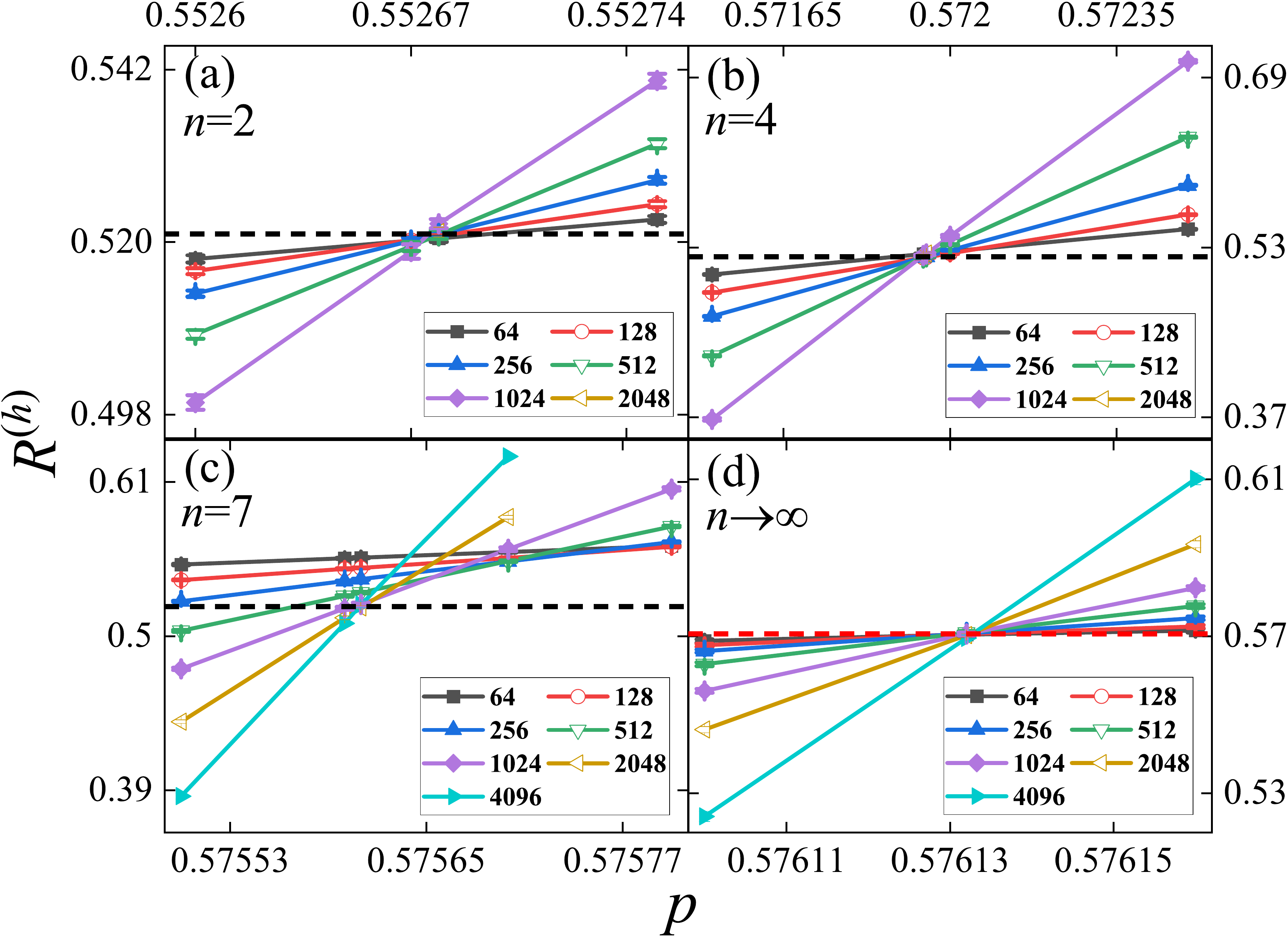}
\caption{Wrapping probability $R^{(h)}$ versus $p$ for the finite generations $n=2$ (a), $4$ (b) and $7$ (c) and for the infinite generation $n \to \infty$ (d) with various sizes. The black dashed lines in panels (a), (b) and (c) represent the exact value $R^{(h)}_c=0.521\,058\,290$ of standard percolation universality, while the red dashed line in panel (d) denotes the estimate $R^{(h)}_c=0.570\,5$ for $n \to \infty$.}~\label{fig_Rh}
\end{figure}

 \textbf{Finite generations.} The finite-size Monte Carlo data of the wrapping probability $R^{(h)}$ are plotted in Figs.~\ref{fig_Rh}(a), (b) and (c) for $n=2$, $4$ and $7$, respectively. For each $n$, the intersections of the $R^{(h)}$ versus $p$ curves for $L \to \infty$ are around a finite bond-occupation probability $p_c$. Meanwhile, the vertical coordinates for the intersections converge to a finite value around $0.521$, which seems universal among these finite generations and agrees with the exact value $R^{(h)}_c=0.521\,058\,290$ of the continuous transition in 2D standard percolation.

By the least-squares criterion, we fit the Monte Carlo data of $R^{(h)}$ to the formula
\begin{eqnarray}\label{fit1}
R^{(h)}(L,p)=R^{(h)}_c&+&a_1 (p-p_c) L^{y_t}+ a_2 (p-p_c)^2 L^{2 y_t}+... \nonumber \\
&+&b_1 L^{-\omega_1} + b_2 L^{-2}+...
\end{eqnarray}
which is an explicit form of (\ref{scalingQs}) for $Q=R^{(h)}$ with additional finite-size correction terms $b_1L^{-\omega_1}$ and $b_2 L^{-2}$. The critical wrapping probability $R^{(h)}_c$ is expected to be universal among the transitions in the same universality and on the same lattice geometry. The parameters $a_k$ and $b_l$ ($k=1,2,...$; $l=1,2,...$) are non-universal. The leading correction exponent is denoted as $\omega_1$, while the subleading correction exponent is fixed to be $-2$.

In the fits, we try to include subleading terms such as $a_2$, $b_1$ and $b_2$ terms or their combinations. This would be useful for a systematical justification on the evidence level of the fits. Besides, preferred fits should feature stability against varying $L_{\rm min}$ that denotes the minimum size incorporated, and ensure that the Chi squared $\chi^2$ per degree of freedom (DF) is not larger than $\scrO(1)$.

For the finite generation $n=2$, we first include the correction terms with $b_1$ and $b_2$, and obtain $p_c=0.552\,679(6)$, $y_t=0.74(4)$ and $R^{(h)}_c=0.522(7)$. These estimates for $y_t$ and $R^{(h)}_c$ further imply a transition in the standard percolation universality. As $R^{(h)}_c$ is fixed to be $R^{(h)}_c=0.521\,058\,290$, we have $p_c=0.552\,678(1)$, $y_t=0.75(4)$ and $\omega_1=1.2(4)$. As $y_t=3/4$ and $\omega_1=1$ are both fixed, we obtain $p_c=0.552\,678(2)$ and $R^{(h)}_c=0.521\,1(4)$. On this basis, we let $y_t=3/4$, $R^{(h)}_c=0.521\,058\,290$ and $\omega_1=1$ all fixed for reducing uncertainties, and obtain $p_c=0.552\,677\,6(9)$.

\begin{table}
\caption{Fits of $R^{(h)}$ to (\ref{fit1}) for the finite generations $n=2$, $4$ and $7$ and for the infinite generation $n \to \infty$.}~\label{tabrh1}
\begin{tabular}{|l|l|l|l|l|l|}
\hline
   $n$ & $p_c$  &  $R^{(h)}_c$   &  $y_t$      & $\omega_1$   & $\chi^2$/DF/$L_{\rm min}$  \\
   \hline
   {\multirow{4}{*}{2}} & 0.552\,679(6) &  0.522(7)   &  0.74(4)        & 0(2)   & 3.7/15/16   \\
   &0.552\,678(1)   & 0.521\,058\,29 &  0.75(4)   & $1.2(4)$    & 3.8/19/8 \\
   &0.552\,678(2)   & 0.521\,1(4) &  3/4   & 1    & 3.7/17/16 \\
   &0.552\,677\,6(9)   & 0.521\,058\,29 &  3/4   & 1   & 3.7/18/16  \\
      \hline
   {\multirow{4}{*}{4}} &0.571\,940(3)   & 0.520(2) &  0.743(6)  & 2(4)   & 7.4/15/64  \\
   &0.571\,941(1)  & 0.521\,058\,29 &  3/4  & 1.0(5)    & 8.8/17/64  \\
   &0.571\,941(2)  & 0.521\,0(8) &  3/4  & 1    & 8.8/17/64  \\
   &0.571\,941(1)  & 0.521\,058\,29 &  3/4  & 1    & 8.6/14/128  \\
      \hline
   {\multirow{3}{*}{7}} &0.575\,613(1)  & 0.521\,058\,29 &  0.77(1)  & 1.2(1)    & 12.3/12/128  \\
   &0.575\,607(4)  &  0.518(3)   & 0.72(5)   & -  &  0.3/2/2048  \\
   &0.575\,608(3)  &   0.519(3)  & 3/4       & -   & 0.7/3/2048   \\
    \hline
      {\multirow{3}{*}{$\infty$}} &0.576\,132\,3(5)  & 0.570\,7(2) & 0.85(2)   & -    & 8.1/11/256  \\
   &0.576\,132\,2(6)  &  0.570\,7(3)   & 0.85(2)   & -  &  7.9/8/512  \\
   &0.576\,131\,7(8)  &   0.570\,2(6)  & 0.86(3)      & -   & 6.7/5/1024   \\
   \hline
\end{tabular}
\end{table}

Similar analyses are performed for $n=4$, for which the results of fits can be found in Table~\ref{tabrh1}.

As indicated by Fig.~\ref{fig_Rh}(c), the finite-size corrections become severe for $n=7$. Hence, simulation results for large lattices are a must to achieve an extensive set of preferred fits. As $R^{(h)}_c=0.521\,058\,290$ is fixed, we find $p_c=0.575\,613(1)$, $y_t=0.77(1)$ and $\omega_1=1.2(1)$. As we incorporate merely large enough sizes with $L_{\rm min}=2048$ and preclude correction terms, the results are $p_c=0.575\,607(4)$, $y_t=0.72(5)$ and $R^{(h)}_c=0.518(3)$. As $y_t=3/4$ is further fixed, we obtain $p_c=0.575\,608(3)$ and $R^{(h)}_c=0.519(3)$.

Finally, by comparing preferred fits, we estimate percolation thresholds as $p_c=0.552\,678(2)$ ($n=2$), $0.571\,941(4)$ ($n=4$) and $0.575\,61(1)$ ($n=7$), where each error bar consists of one statistical error and a subjective estimate of systematic error. The estimates of critical wrapping probabilities are $R^{(h)}_c=0.521\,2(6)$ ($n=2$), $0.521(2)$ ($n=4$) and $0.519(4)$ ($n=7$), which agree well with the exact value $R^{(h)}_c=0.521\,058\,290$ of 2D standard percolation. Moreover, the correction term with $L^{-\omega_1}$ ($\omega_1 \approx 1.2$) emerges in FSS, differing from the leading correction term with $L^{-2}$ of critical wrapping probabilities in standard percolation~\cite{newman2001fast}. It is noteworthy that the correction exponent $\omega_1=3/2$ was also found for 2D percolation~\cite{Ziff2011Corr}.

\textbf{Infinite generation.} For the infinite generation $n \to \infty$, the $R^{(h)}$ versus $p$ curves are plotted in Fig.~\ref{fig_Rh}(d), which demonstrates that the intersections are located around $p_c \approx 0.576\,13$, where $R^{(h)}$ is close to $R^{(h)}_c \approx 0.570$. The intersections are nearly coincident at (0.576\,13, 0.570), indicating a continuous transition and minor finite-size corrections. Hence, we perform fits to (\ref{fit1}) by dropping correction terms. As shown in Table~\ref{tabrh1}, the results obtained from the fits with $L_{\rm min}=256$, $512$ and $1024$ are consistent. The final estimates of percolation threshold and critical wrapping probability are $p_c=0.576\,132(2)$ and $R^{(h)}_c=0.570\,5(8)$, respectively. The value of $R^{(h)}_c$ differs from that of standard percolation and indicates a new universality class.

\begin{figure}
\centering
\includegraphics[width=8cm,height=8cm]{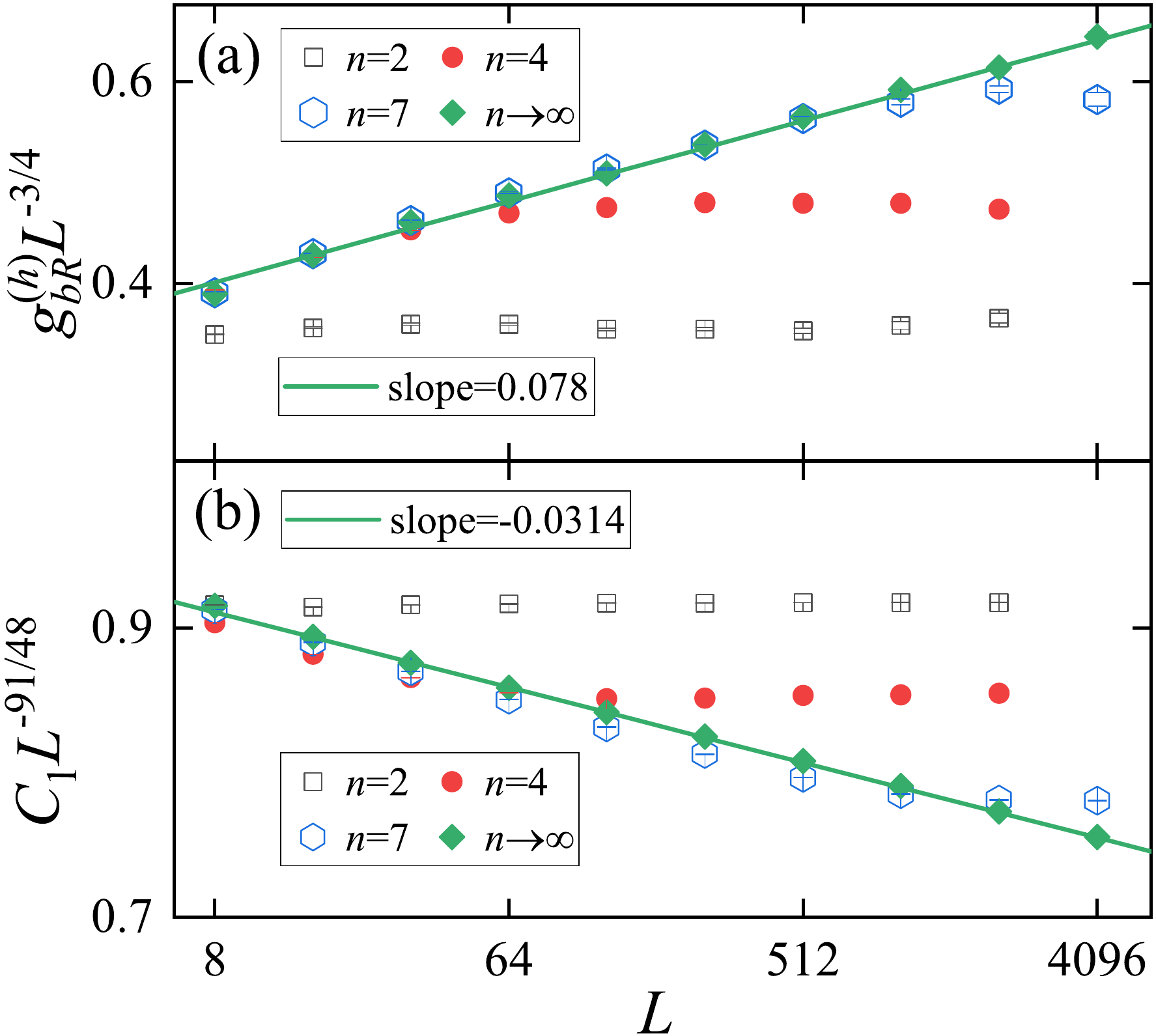}
\caption{Illustration for the estimated critical exponents $y_t$ and $y_h$ by scaled quantities $g_{bR}^{(h)}L^{-3/4}$ (a) and $C_1L^{-91/48}$ (b). For the finite generations $n=2$, $4$ and $7$, the scaled data are asymptotically constants as $L$ increases, confirming the estimates of the critical exponents as $y_t=3/4$ and $y_h=91/48$. For the infinite generation $n \to \infty$, deviations from the behavior of standard percolation universality are indicated by the linearities of scaled data with non-zero slopes. These slopes take values $0.078$ and $-0.0314$, which measure the deviations from $y_t=0.828(5)$ and $y_h=1.864\,4(7)$ to $y_t=3/4$ and $y_h=91/48$, respectively.}~\label{figgC1}
\end{figure}

\subsection{Critical exponents $y_t$ and $y_h$}\label{cpyy}

We now focus on the Monte Carlo data at the above-estimated percolation thresholds, where the FSS formula (\ref{scalingQs}) is simplified as
\begin{equation}\label{fssO}
Q(L,p_c)=L^{X_Q}(a_0+b_1 L^{-1} + b_2 L^{-2}+...)
\end{equation}
with finite-size correction terms $b_1$ and $b_2$, and the constant $a_0$. In some cases, a constant term $c_0$ from analytic background should be included in addition to (\ref{fssO}).

\textbf{The critical exponent $y_t$.}  We estimate the critical exponent $y_t$ from the covariance $g^{(h)}_{bR}$ which relates to the derivative ${dR^{(h)}}/{dp}$.
Fits are performed according to (\ref{fssO}) with $X_Q=y_t$. For $n=2$, we obtain $y_t=0.751(3)$ with $\chi^2/{\rm DF}=1.3/2$ and $L_{\rm min}=128$, as the constant term $c_0$ and the correction terms $b_1$ and $b_2$ are not included. Similarly, for $n=4$, we obtain $y_t=0.747(3)$ with $\chi^2/{\rm DF}=1.1/2$ and $L_{\rm min}=256$. For $n=7$, stable fits are achieved if $c_0$ term is present. Accordingly, we have $y_t=0.76(1)$ with $\chi^2/{\rm DF}=0.1/1$ and $L_{\rm min}=256$. The final estimate of $y_t$ for each of the finite generations is achieved by comparing preferred fits. As summarized in Table~\ref{tab1}, the results of $y_t$ consist with the exact value $y_t=3/4$ of standard 2D percolation. For the infinite generation $n \to \infty$, as listed in Table~\ref{tab2}, we obtain $y_t=0.827(1)$, $0.829(1)$ and $0.827(2)$, with $L_{\rm min}=64$, $128$ and $256$, respectively. As given in Table~\ref{tab4}, our final estimate of $y_t$ for $n \to \infty$ is $y_t=0.828(5)$; the error bar is enlarged, since some finite-size corrections might be ignored in the fitting formula.

For illustrating $y_t$, we plot in Fig.~\ref{figgC1}(a) the scaled covariance $g^{(h)}_{bR} L^{-{3/4}}$ for various generations. For the finite generations $n=2$, $4$ and $7$, the scaled data eventually become constants as $L$ increases. For the infinite generation $n \to \infty$, deviation from the behavior of standard percolation is demonstrated by the non-zero slope $0.078$, which measures the distance from $y_t=0.828(5)$ to $y_t=3/4$.

\begin{figure}
\centering
\includegraphics[width=8.8cm,height=7.5cm]{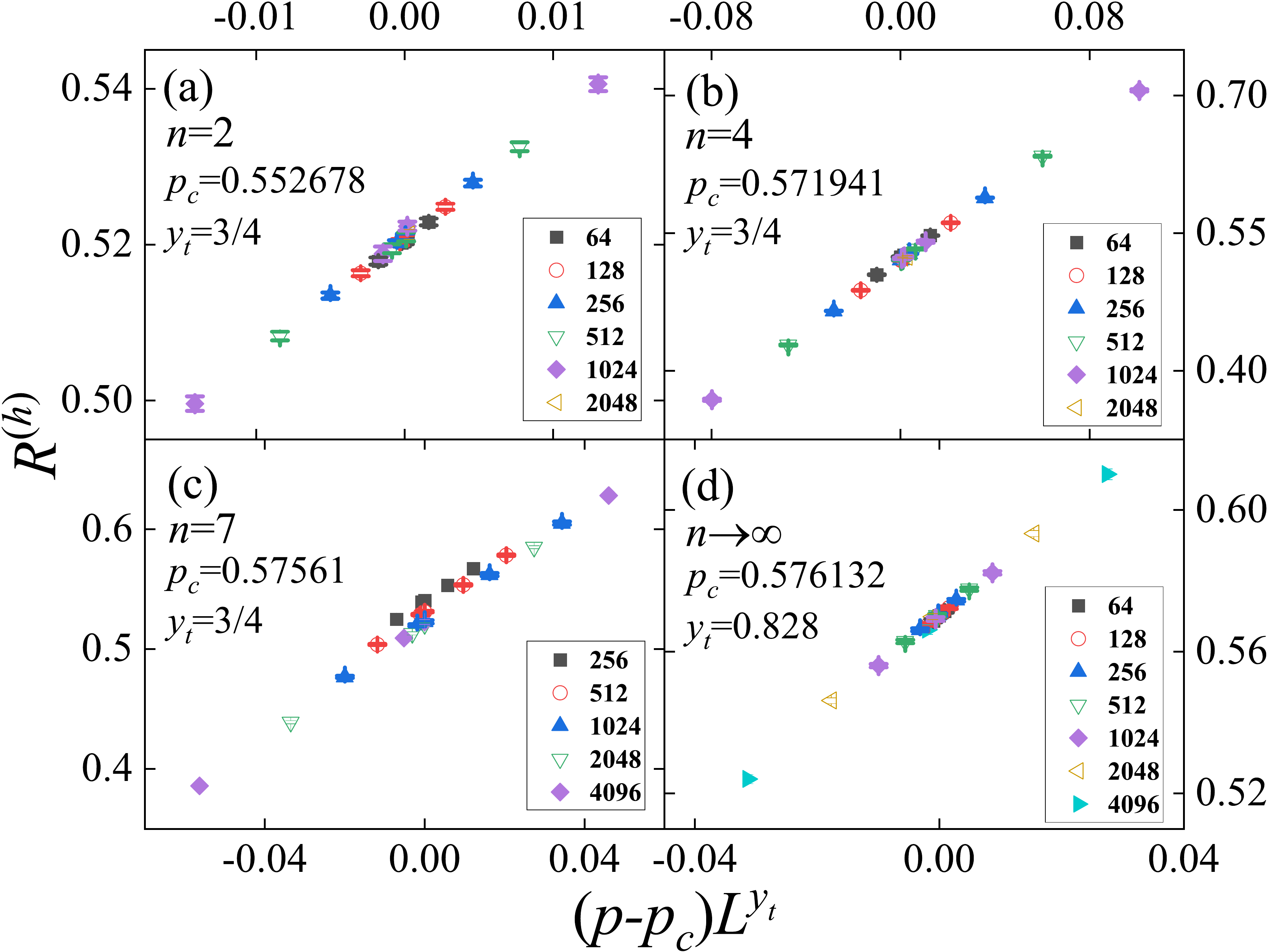}
\caption{Illustration for the scaling formula (\ref{scalingQs}) with $Q=R^{(h)}$. In panels (a), (b) and (c), the $R^{(h)}$ data for finite generations are plotted against $(p-p_c)L^{y_t}$ with $y_t=3/4$ and $p_c=0.552\,678$ ($n=2$), $0.571\,941$ ($n=4$) and $0.575\,61$ ($n=7$). In panel (d), the $R^{(h)}$ data for the infinite generation $n \to \infty$ are plotted with $y_t=0.828$ and $p_c=0.576\,132$.}~\label{fig_Rh2}
\end{figure}

Another verification for the estimated $y_t$ is provided by Fig.~\ref{fig_Rh2}, where we plot $R^{(h)}$ versus $(p-p_c)L^{y_t}$ for various $n$. These plots serve as simultaneous illustrations for the estimated $y_t$ ($3/4$ for $n=2$, $4$, $7$ and 0.828 for $n \to \infty$), the estimated $p_c$, and the scaling formula (\ref{scalingQs}). For each $n$, the scaled data of various $L$ collapse compactly on top of each other as $L \to \infty$.

\textbf{The critical exponent $y_h$.}  The critical exponent $y_h$ can be estimated from $C_1$ and $\chi$ according to (\ref{fssO}) with $X_Q=y_h$ and $2y_h-2$, respectively. For $n=2$, we obtain $y_h=1.895\,9(2)$ by $C_1$, and $2y_h-2=1.792\,1(4)$ by $\chi$. For $n=4$, we have $y_h=1.896\,2(5)$ by $C_1$, and $2y_h-2= 1.792(1)$ by $\chi$. For $n=7$, we obtain $y_h=1.894\,8(9)$ by $C_1$, and $2y_h-2=1.792(2)$ by $\chi$. The final estimates of $y_h$ for finite generations are given as $1.896\,0(3)$ ($n=2$), $1.896\,5(8)$ ($n=4$) and $1.895(2)$ ($n=7$), which are consistent with the exact value $y_h=91/48$ of standard percolation. The results for the infinite generation $n \to \infty$ are exemplified in Table~\ref{tab2}, and our final estimate of $y_h$ is $y_h=1.864\,4(7)$.

In Fig.~\ref{figgC1}(b), the scaled data $C_1 L^{-91/48}$ for finite generations converge to constants as $L$ increases, confirming $y_h=91/48$. Meanwhile, the scaled data for $n \to \infty$ clearly deviate from the behavior in standard percolation universality and confirm $y_h=1.864\,4(7)$.

\begin{table}
\caption{Final estimates of the percolation thresholds $p_c$, the critical exponents $y_t$ and $y_h$, and the critical wrapping probabilities $R^{(h)}_c$ for the finite generations $n$=$2$, $4$ and $7$. Exact values in the 2D standard universality class are listed as well for comparison.}~\label{tab1}
\begin{tabular}{|c|c|c|c|c|}
\hline
   $n$ & 2 &  4 &  7  & exact  \\
\hline
  $p_c$  & 0.552\,678(2) & 0.571\,941(4) & 0.575\,61(1) & - \\
  $y_t$  & 0.751(5) &  0.74(1) & 0.75(2)  & 3/4 \\
  $y_h$  & 1.896\,0(3) & 1.896\,5(8) & 1.895(2) & 91/48 \\
  $R^{(h)}_c$ & 0.521\,2(6) & 0.521(2)  & 0.519(4) & 0.521\,058\,290\\
 \hline
\end{tabular}
\end{table}

\begin{table}
\caption{Fits of $g^{(h)}_{bR}$, $C_1$ and $\chi$ to (\ref{fssO}) for the infinite generation $n \to \infty$. The scaling exponents $X_Q$ for $g^{(h)}_{bR}$, $C_1$ and $\chi$ are $y_t$, $y_h$ and $2y_h-2$, respectively. Our final estimates $y_t=0.828(5)$ and $y_h=1.864\,4(7)$ are based on comparing all preferred fits of these quantities.}~\label{tab2}
\begin{tabular}{|l|l|l|}
\cline{2-3}
\multicolumn{1}{c}{\quad} &
 \multicolumn{2}{|c|}{$n \to \infty$}  \\
 \hline
   $Q$  &  $X_Q$  &  $\chi^2/{\rm DF}/L_{\min}$   \\
  \hline
  {\multirow{3}{*}{$g^{(h)}_{bR}$}} & 0.827(1) &  11.0/5/64 \\
  & 0.829(1) &  3.9/4/128 \\
  & 0.827(2) &  2.5/3/256 \\
  \hline
  {\multirow{3}{*}{$C_1$}} & 1.864\,7(1) & 14.2/5/32 \\
  & 1.864\,3(2) & 4.6/4/64  \\
  & 1.864\,0(3) & 2.1/3/128  \\
  \hline
   {\multirow{3}{*}{$\chi$}} & 1.729\,1(1) & 7.4/5/64 \\
   & 1.729\,3(1) & 5.4/4/128 \\
   & 1.729\,1(2) & 4.5/3/256 \\
\hline
\end{tabular}
\end{table}

\begin{table}
\caption{Final estimates of the percolation threshold $p_c$, the critical exponents $y_t$ and $y_h$, and the critical wrapping probability $R^{(h)}_c$ for the infinite generation $n \to \infty$.}~\label{tab4}
\begin{tabular}{|l|l|l|l|l|}
\hline
  $n$ & $p_c$ &  $y_t$ & $y_h$ & $R^{(h)}_c$ \\
   \hline
   $\infty$ & 0.576\,132(2)&  0.828(5)&  1.864\,4(7) &  0.570\,5(8)   \\
 \hline
\end{tabular}
\end{table}

\begin{figure}
\centering
\includegraphics[width=8cm,height=10cm]{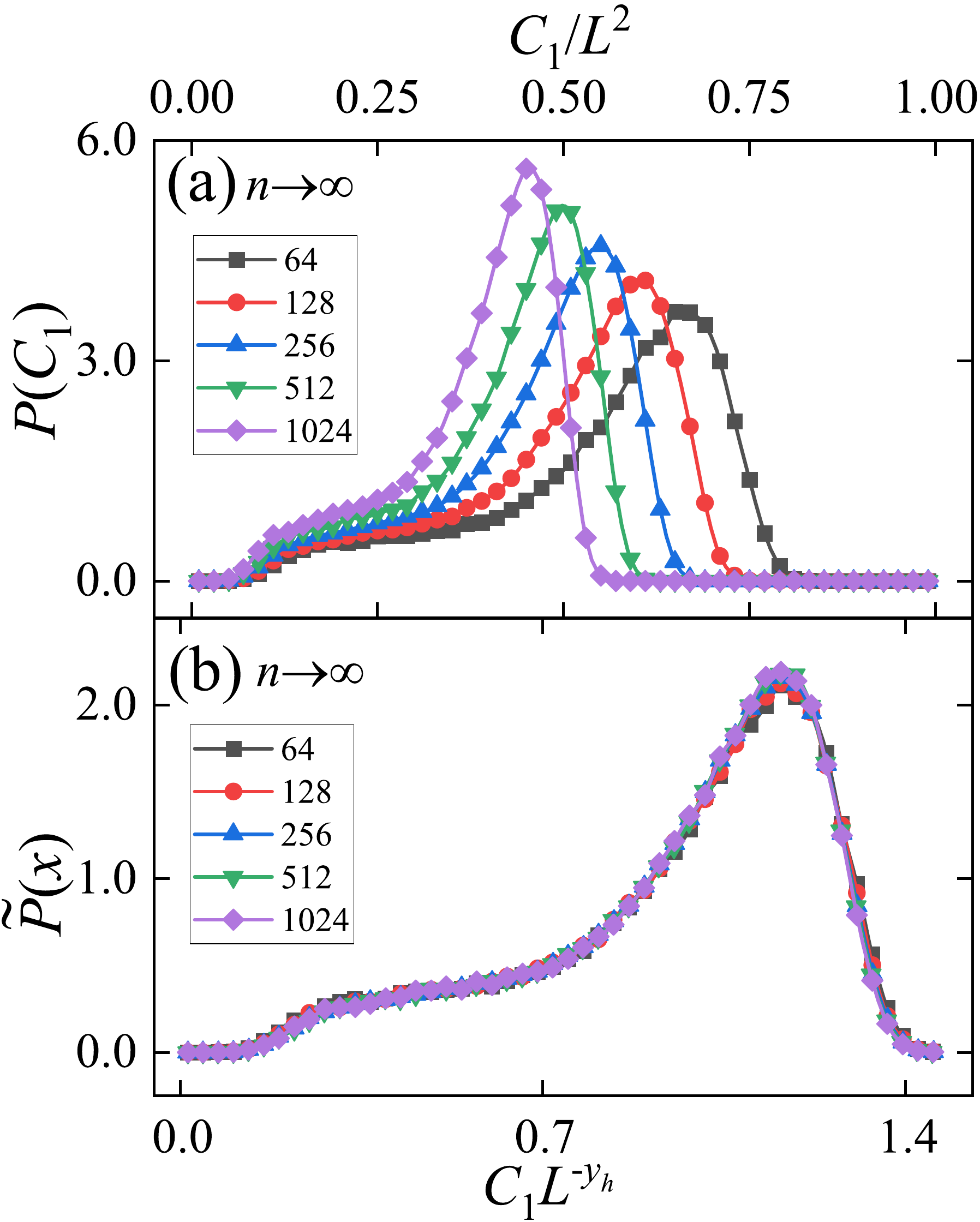}
\caption{Critical distribution function $P(\scrC_1,L)$ for the scaled size $\scrC_1 L^{-2}$ of the largest cluster in the infinite generation $n \to \infty$. The data are for $p_c=0.576\,132$. In panel (b), the rescaled critical distribution function $\tilde{P}(x)$ ($x \equiv \scrC_1/{L^{y_h}}$) shows a single-variable behavior.}~\label{fig_his_C1}
\end{figure}

\begin{figure}
\centering
\includegraphics[width=8cm,height=10cm]{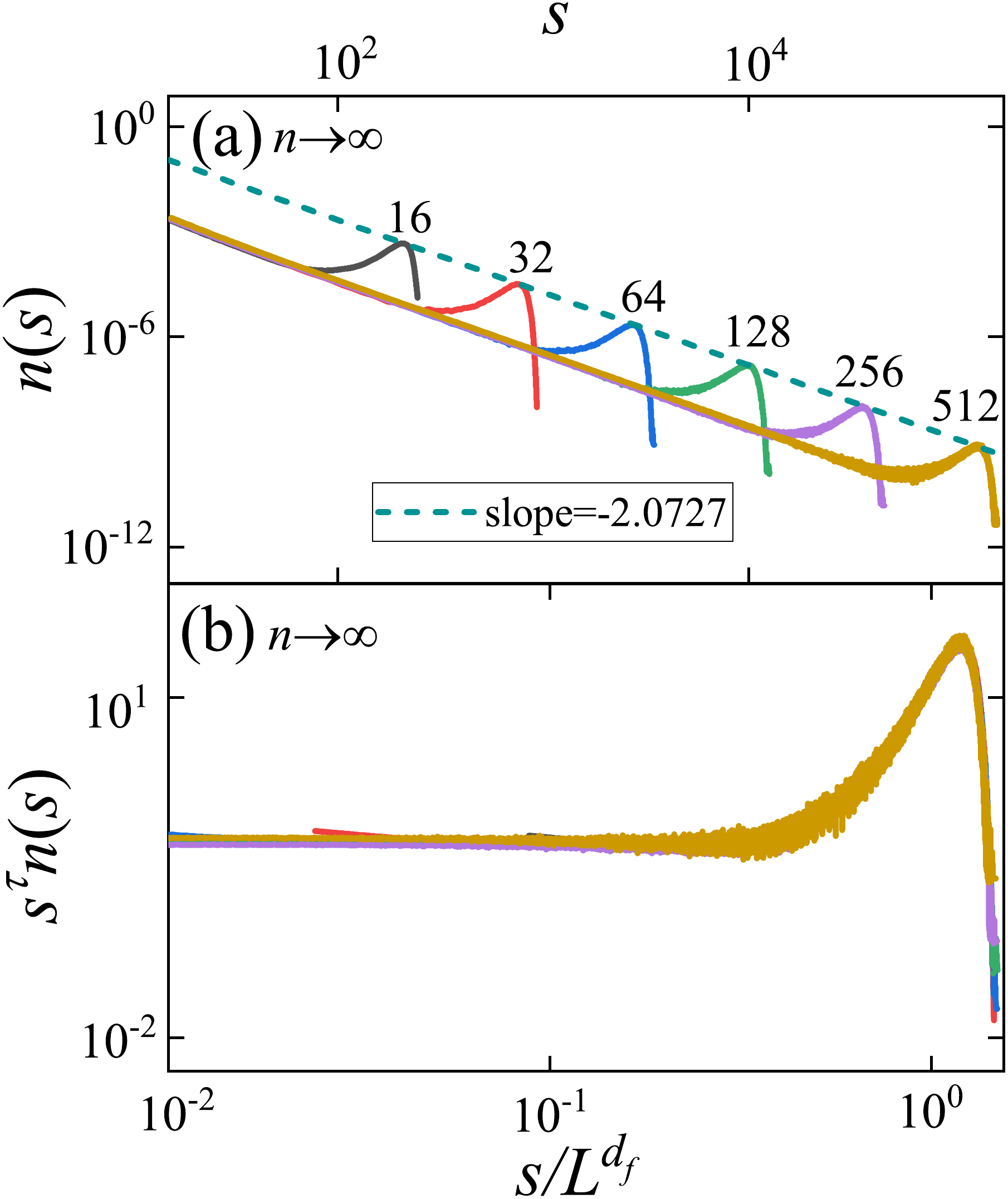}
\caption{Cluster-number density $n(s,L)$ at the percolation threshold $p_c=0.576\,132$ of the infinite generation $n \to \infty$, with $L=16, 32, 64, 128, 256$ and $512$. In panel (a), the dashed line has a slope of $-\tau$ with $\tau=1+2/d_f \approx 2.072\,7$. A scaling analysis according to (\ref{nsscaling}) is shown in panel (b).}~\label{fignsscs}
\end{figure}

\subsection{Geometric properties of critical clusters}\label{csci}

\begin{figure}
\centering
\includegraphics[width=8cm,height=7cm]{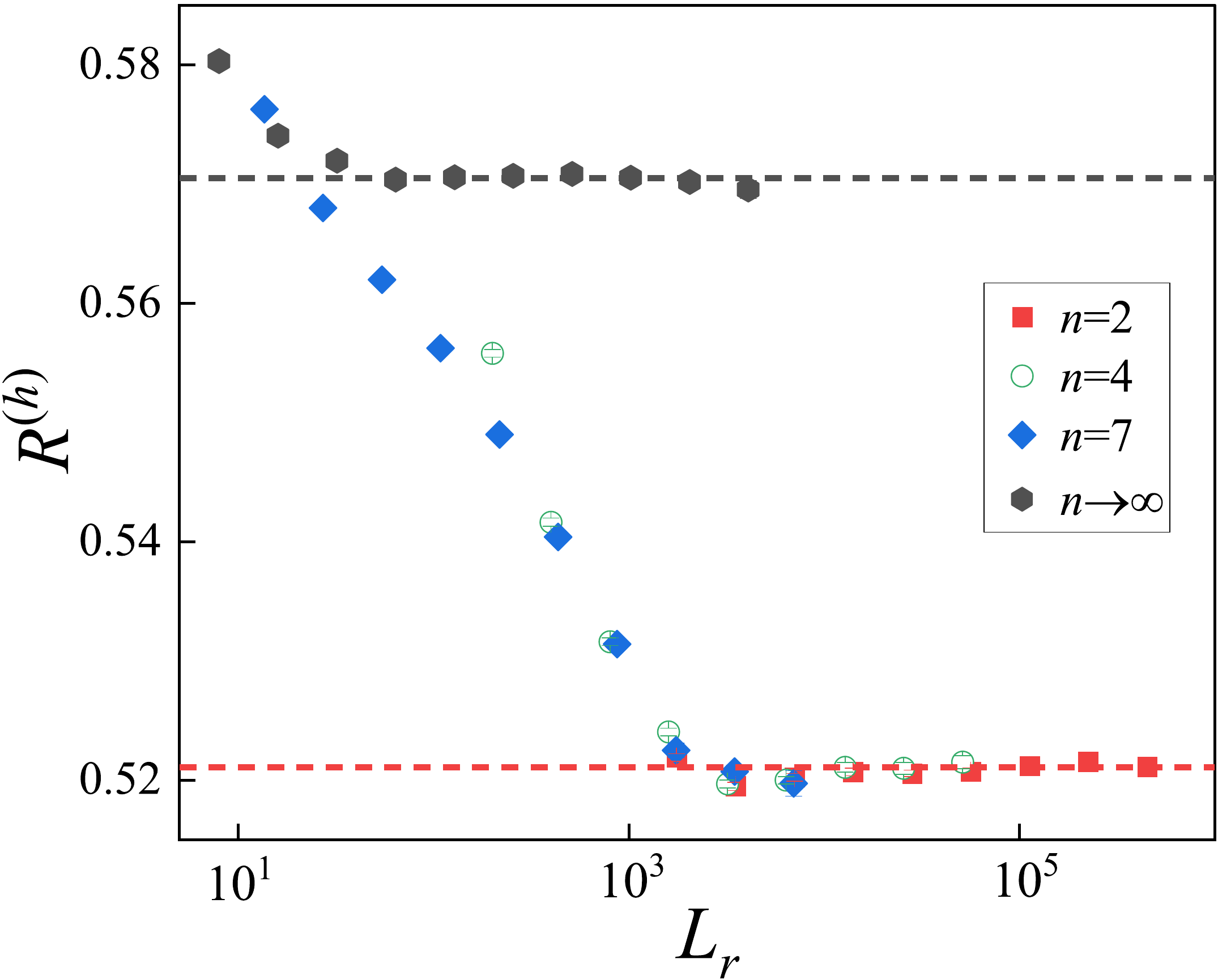}
\caption{Finite-size $R^{(h)}$ data at $p_c$ versus rescaled size $L_r=L/r(n)$ for finite and infinite generations. The rescaled factor
 $r(n)$ is generation-dependent. The asymptotic values (represented by dashed lines) in the $L_r  \to \infty$ limit are $0.570\,5(8)$ and $0.521\,058\,290$.}~\label{figRhcr}
\end{figure}

\newcommand{\tabincell}[2]{\begin{tabular}{@{}#1@{}}#2\end{tabular}}

\begin{figure}
\centering
\includegraphics[width=8cm,height=7cm]{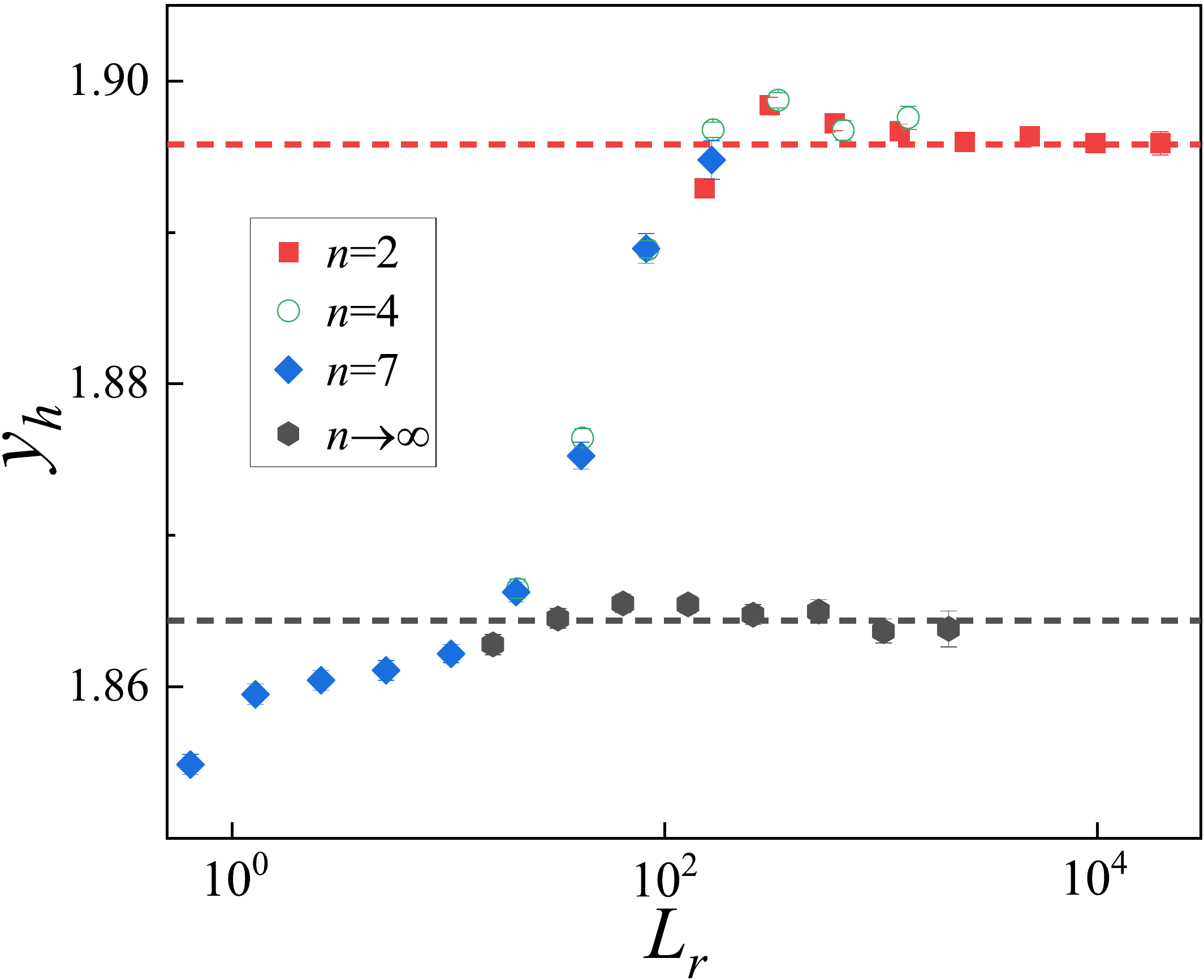}
\caption{Finite-size estimates of the effective magnetic exponent $y_h$, determined from $C_1$ at $p_c$, versus rescaled size $L_r$. The asymptotic values (represented by dashed lines) in the $L_r \to \infty$ limit are $1.864\,4(7)$ and $91/48$.}~\label{figyhcr}
\end{figure}

We have found that the transition in infinite generation is continuous and falls outside the universality of standard percolation. In following, we explore the geometric properties of critical clusters. We investigate the critical probability distribution of largest-cluster size as well as the critical cluster-number density, and examine their compatibility with the FSS theory of continuous geometric transition.

The critical distribution function $P(\scrC_1,L)$ for the largest-cluster size $\scrC_1$ is shown in Fig.~\ref{fig_his_C1}(a). At the critical bond-occupation probability $p_c=0.576\,132$ (shown in the plot) and  its neighborhood, we do not find a stable double-peaked structure, confirming the absence of discontinuous transition. Further, as displayed in Fig.~\ref{fig_his_C1}(b), the distribution $P(\scrC_1,L) d \scrC_1$ can be rescaled into a single-variable form as $\tilde{P}(x)dx$ with $x \equiv \scrC_1/{L^{d_f}}$ and $d_f=y_h$. The absence of double-peaked structure and the single-variable behavior in distribution function are indicators for a continuous geometric transition.

We analyze the cluster-number density $n(s,L)$ of cluster size $s$. At a continuous phase transition, it is expected that
\begin{equation}\label{nsscaling}
n(s,L)=s^{-\tau} \tilde{n}(s/L^{d_f}),
\end{equation}
where $\tilde{n}$ is a scaling function and $\tau=1+2/d_f$. As shown in Fig.~\ref{fignsscs}(a), the large-$s$ asymptotics of $n(s,L)$ is consistent with the power law $s^{-\tau}$, where the exponent $\tau=2.072\,7$ relates to $d_f=y_h=1.864\,4$. Figure~\ref{fignsscs}(b) plots $s^{\tau}n(s,L)$ versus $s/L^{d_f}$ for various $L$ and demonstrates a compact collapse. We hence conclude that the FSS formula (\ref{nsscaling}) is compatible with the percolation transition in infinite generation.

\subsection{Continuous crossover}\label{ccifg}

As displayed in Fig.~\ref{figgC1}, the quantities $g^{(h)}_{bR}$ and $C_1$ exhibit more severe finite-size corrections as the finite $n$ increases. A continuous crossover of critical behavior from infinite to finite generation is indicated, as the $n=7$ and $n \to \infty$ data are close at small sizes but deviate at larger sizes. In following, we further illustrate the crossover phenomenon.

Figure~\ref{figRhcr} displays a collapse of the critical wrapping probabilities $R^{(h)}(L,p_c)$ versus an $n$-dependent rescaled size $L_r$. We define $L_r=L/r(n)$ with the rescaled factor $r(n)$ chosen such that the $R^{(h)}$ versus $L_r$ data of various $n$ collapse on top of each other. As a result, the small-size data for $n=7$ are close to those for $n \to \infty$. By contrast, the $n=7$ data at large $L_r$ tend to collapse with the data for $n=2$ and $4$.

The crossover can also be seen in the size-dependent effective magnetic exponent $y_h(L)$, which is determined from $C_1$ at sizes $L$ and $2L$ by the formula $y_h(L)={\rm ln}[C_1(2L,p_c)/C_1(L,p_c)]/{\rm ln}2$. The results are shown in Fig.~\ref{figyhcr}, which demonstrates a crossover between the asymptotics of $n=7$ data. At small sizes the $n=7$ data are in the same profile with $n \to \infty$, whereas at large sizes they collapse with the data of $n=2$ and $4$. Moreover, we observe the crossover phenomenon by the effective thermal exponent $y_t(L)$ extracted from $g^{(h)}_{bR}$ (not shown), although it suffers from huge statistical errors.

\section{Discussion}\label{sm}
By using extensive Monte Carlo simulations, we study the critical behavior of the HDP on the square lattice. At the finite generations $n=2$, $4$ and $7$, we locate percolation transitions at $p_c=0.552\,678(2)$, $0.571\,941(4)$ and $0.575\,61(1)$, respectively. We find that these transitions belong to the universality class of standard percolation in two dimensions, although finite-size corrections become larger when $n$ increases, as demonstrated in Figs.~\ref{fig_Rh}, \ref{figgC1} and \ref{fig_Rh2}. At the infinite generation $n \to \infty$, we observe a continuous transition at $p_c=0.576\,132(2)$ with the critical exponents $y_t=0.828(5)$ and $y_h=1.864\,4(7)$ and the critical wrapping probability $R^{(h)}_c=0.570\,5(8)$, which fall outside the standard percolation universality. The critical distribution function $P(\scrC_1,L)$ and the critical cluster-number density $n(s,L)$ follow the standard scaling behavior of a continuous geometric transition.

\begin{figure}
\centering
\includegraphics[width=8cm,height=6cm]{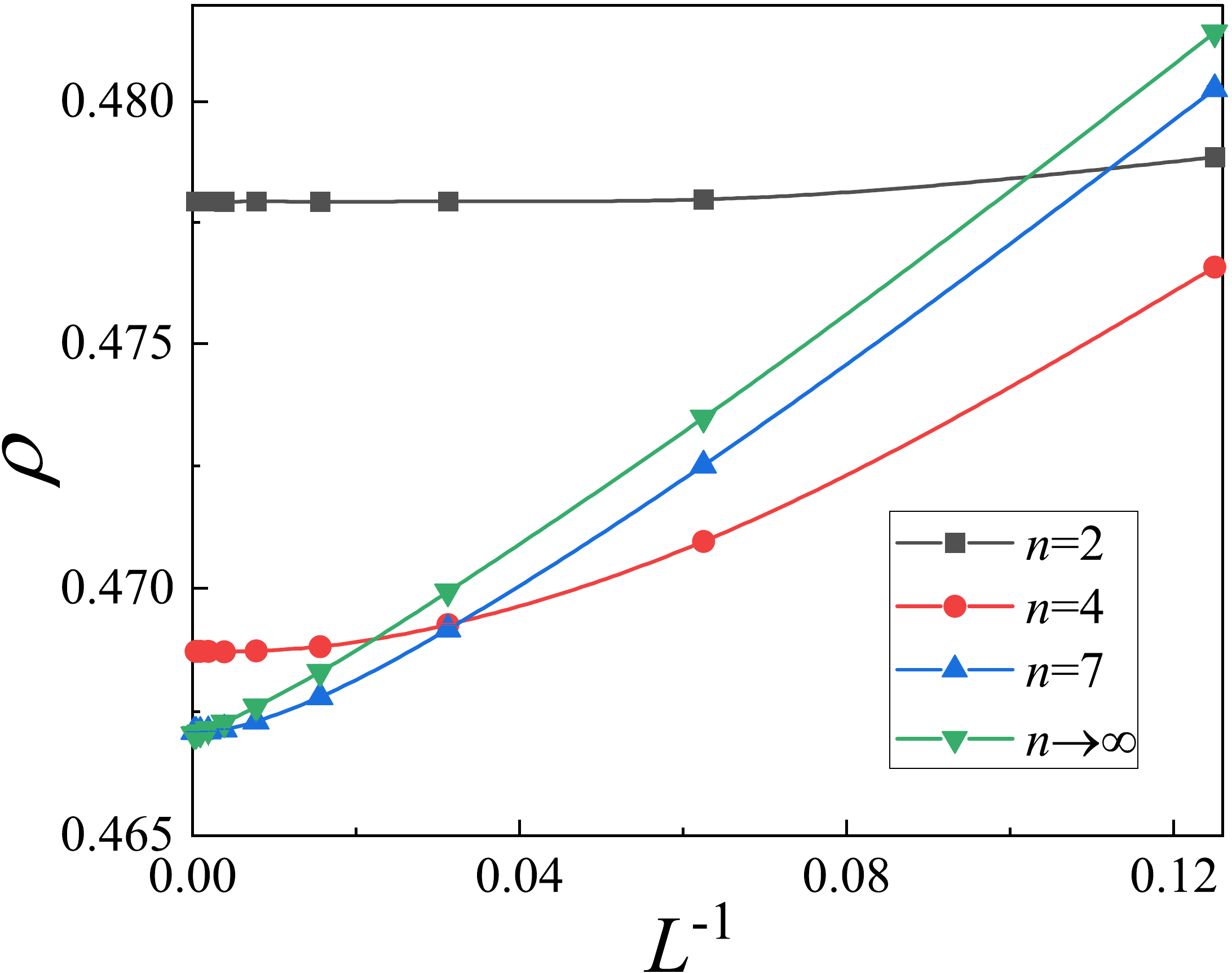}
\caption{Occupied-bond density $\rho$ versus $L^{-1}$ at the percolation thresholds of finite and infinite generations. Simulations are performed at the estimated percolation thresholds $p_c=0.552\,678$ ($n=2$), $0.571\,941$ ($n=4$), $0.575\,61$ ($n=7$) and $0.576\,132$ ($n \to \infty$). Critical densities of occupied bonds are obtained as $\rho_c=0.477\,9(1)$ ($n=2$), $0.468\,7(1)$ ($n=4$), $0.467\,1(4)$ ($n=7$) and $0.467\,0(7)$ ($n \to \infty$).}~\label{fig_effective_pc}
\end{figure}

As shown in Fig.~\ref{fig_np_n}, $n_{\rm sat}$ reaches the maximum at $p=0.576\,132$. As we have known, the bond-occupation probability $p =0.576\,132$ is the percolation threshold of the infinite generation. Hence, at $p=0.576\,132$, initial percolation configurations typically have the largest number of iterative bond deletions before reaching the infinite-generation configurations, which are critical.

We obtain complementary evidence confirming a continuous crossover of critical behavior from infinite to finite generation. The evidence comes from various quantities such as $g^{(h)}_{bR}$ (Fig.~\ref{figgC1}(a)), $C_1$ (Fig.~\ref{figgC1}(b)) and $R^{(h)}$ (Fig.~\ref{figRhcr}) and from effective critical exponents (Fig.~\ref{figyhcr}). Here, we give more-detailed descriptions. Notice that $n_{\rm sat}$ diverges with $L$ in the parameter regime of interest (Fig.~\ref{fig_np_n}). Hence, for a finite generation, the value of $n_{\rm sat}$ at small $L$ may be smaller than or comparable with $n$, and the system behaves {\it like} infinite generation due to finite-size effects. By contrast, as $L$ increases, $n_{\rm sat}$ ultimately exceeds the finite $n$, and the intrinsic finite-generation behavior is recovered.

Recall that $\rho$ is the density of occupied bonds remaining at a given generation. Figure~\ref{fig_effective_pc} shows the $L$ dependence of $\rho$ at the percolation thresholds of various generations. For each generation, let the asymptotic critical value be $\rho_c = \rho(p=p_c, L \rightarrow \infty)$. We obtain the critical occupied-bond densities as $\rho_c=0.477\,9(1)$ ($n=2$), $0.468\,7(1)$ ($n=4$), $0.467\,1(4)$ ($n=7$) and $0.467\,0(7)$ ($n \to \infty$). In contrast to the $p_c-n$ dependence, the critical density $\rho_c$ decreases with increasing $n$.

There are a number of open questions motivated by this work. Some of them are as follows. What are the exact values of the critical exponents $y_t=0.828(5)$ and $y_h=1.864\,4(7)$ for the infinite generation? Can they be obtained within the framework of conformal field theories or Coulomb gas theory? It is noted that the HDP can be extended such that it has an arbitrary number $N_{\ell}$ of layers. In the present $N_{\ell}=2$ model shown in Fig.~\ref{fig_guide}, the inter-layer coupling looks like A-B-A'-B'-..., and thus has an effective period of 2. For $N_{\ell} \geq 3$, the configuration coupling looks like A-B-C...-A'-B'-C'..., and the coupling period is longer. Specific treatment can be that a bond gets deleted if the sites are in different clusters in all the intermediate layers or in any of the intermediate layers. As $N_{\ell}$ is sufficiently large, percolation clusters in the infinite-generation limit are either very dense or very small to resist the bond-deletion action. A first-order percolation transition may arise for some large $N_{\ell}$ and finite generation $n$. An intriguing scenario then can happen: for a given large $N_{\ell}$, there exists a ``tricritical" value of $n$, separating a line of continuous and first-order percolation transitions. The randomly networked structure would be a key factor for the origin of first-order percolation transition. More understanding can arise from exploring the HDP in higher spatial dimensions $d \geq 3$ and on the complete graph. We leave these questions for further studies.

\begin{acknowledgments}
\textit{Acknowledgements.} YD is indebted to Ming Li and Linyuan L\"{u} for valuable discussions. This work has been supported by the National Natural Science Foundation of China under Grant Nos.~11774002, 11625522 and 11975024, and the Department of Education in Anhui Province.
\end{acknowledgments}

\bibliography{papers}

\end{document}